\documentclass[prd,letterpaper,twocolumn,preprintnumbers,nofootinbib]{revtex4}

\usepackage{slashed}
\usepackage{amsmath,amssymb}
\usepackage{graphicx}
\usepackage{units}
\usepackage{bbold}
\usepackage{xcolor}
\usepackage{dsfont}
\usepackage[hyperfootnotes=false,colorlinks,citecolor=blue]{hyperref}

\usepackage{comment}
\usepackage[normalem]{ulem}     

\newcommand{\beq}{\begin{equation}}
\newcommand{\eeq}{\end{equation}}
\newcommand{\bea}{\begin{eqnarray}}
\newcommand{\ena}{\end{eqnarray}}
\newcommand{\dd}{{\rm d}}
\newcommand{\ii}{{\rm i}}

\def \epsilon {\varepsilon} 

\def \vec#1{{\boldsymbol{#1}}}

\newcommand{\avb}[1]{\big\langle #1 \big\rangle}	
\newcommand{\eq}{\mathrm{eq}}
\newcommand{\ie}{\emph{i.e.}}
\newcommand{\eg}{\emph{e.g.}}

\newcommand{\matrixx}[1]{\begin{pmatrix} #1 \end{pmatrix}} 

\newcommand{\Hn}{H}
\newcommand{\An}{A}
\newcommand{\Hpm}{{H^+}}
\newcommand{\YID}{Y_{\eta}}

\begin{document}

\title{Dark matter and radiative neutrino masses in conversion-driven scotogenesis}

\author{Julian Heeck}
\email[E-mail: ]{heeck@virginia.edu}
\thanks{ORCID: \href{https://orcid.org/0000-0003-2653-5962}{0000-0003-2653-5962}.}
\affiliation{Department of Physics, University of Virginia,
Charlottesville, Virginia 22904-4714, USA}

\author{Jan Heisig}
\email[E-mail: ]{heisig@virginia.edu}
\affiliation{Department of Physics, University of Virginia,
Charlottesville, Virginia 22904-4714, USA}

\author{Anil Thapa}
\email[E-mail: ]{wtd8kz@virginia.edu}
\thanks{ORCID: \href{https://orcid.org/0000-0003-4471-2336}{0000-0003-4471-2336}.}
\affiliation{Department of Physics, University of Virginia,
Charlottesville, Virginia 22904-4714, USA}

\begin{abstract}
The scotogenic model generates Majorana neutrino masses radiatively, with dark matter particles running in the loop. We explore the parameter space in which the relic density of fermionic dark matter is generated via a conversion-driven freeze-out mechanism. The necessity for small Yukawa couplings to initiate chemical decoupling for conversion processes naturally reproduces small neutrino masses as long as the active neutrinos are hierarchical. The model can also resolve the recently reported deviation in the $W$-boson mass while satisfying constraints from direct detection, charged lepton flavor violation as well as collider bounds. Parts of the parameter space lead to long-lived particle signatures to be probed at the LHC.
\end{abstract}

\maketitle

\section{Introduction}
\label{sec:intro}

The scotogenic model~\cite{Ma:2006km} is a simple explanation of neutrino masses and dark matter (DM), connecting these two strongest proofs for physics beyond the Standard Model (SM).
As a particularly welcome feature, neutrino masses generated in the scotogenic model are naturally small due to loop suppression factors and the heavy DM mass. DM consists of either the lightest new fermion or the lightest neutral scalar; the latter is highly reminiscent of the inert-doublet model~\cite{Deshpande:1977rw,Barbieri:2006dq,LopezHonorez:2006gr}.  Constraints from direct and indirect detection as well as charged lepton flavor violation restrict the parameter space and offer possibilities of verifying this model (see Refs.~\cite{Borah:2020wut,deBoer:2021pon,Liu:2022byu} for recent work in this direction).

The loop diagrams leading to neutrino masses contain DM and, thus, link neutrino parameters to DM couplings. These couplings are severely constrained in the region of parameter space where DM is produced via thermal freeze-out.
However, our incomplete knowledge of the active-neutrino parameters, notably the mass of the lightest neutrino, makes it possible to suppress some of the DM Yukawa couplings to an arbitrary degree. This might be considered fine-tuning, but opens up the phenomenologically interesting region of \emph{freeze-in} DM~\cite{McDonald:2001vt,Asaka:2005cn,Hall:2009bx} explored in Refs.~\cite{Molinaro:2014lfa,Hessler:2016kwm,Baumholzer:2019twf} within the scotogenic model.

In this article, we point out that there is a region of parameter space in which the Yukawa couplings are between the freeze-in and freeze-out values,  enabling a \emph{conversion-driven}~\cite{Garny:2017rxs} (or coscattering~\cite{DAgnolo:2017dbv}) freeze-out scenario. No particular fine-tuning is required, all entries of the Yukawa matrix can even be of similar order of magnitude. The moderately small Yukawas necessary for this new viable region have implications for the lightest active neutrino mass and naturally suppress any direct and indirect detection signatures, as well as charged lepton flavor violation.

Motivated by the recent precision measurement of the $W$-boson mass by CDF~\cite{CDF:2022hxs}, which exceeds the SM prediction by $7\sigma$, we focus our analysis on the region of parameter space that can alleviate this tension, although this is not strictly necessary for successful conversion-driven freeze-out. The key ingredient for a larger $W$-boson mass is the mass splitting within the new scalar $SU(2)_L$ doublet, which propagates to the $W$-boson mass at the one-loop level. This solution is, of course, identical to many other two-Higgs-doublet solutions of the CDF anomaly, but is connected to both neutrino mass and DM phenomenology in the scotogenic model and, hence, less ad hoc. 
Related studies of the impact of the CDF measurement on the scotogenic  and inert-doublet models have already been presented in Ref.~\cite{Batra:2022pej} and Refs.~\cite{Fan:2022dck,Zhu:2022tpr}, respectively, but with focus on different regions of parameter space.

The rest of this article is organized as follows:
We introduce the scotogenic model and our notation in Sec.~\ref{sec:model}. In Sec.~\ref{sec:W-boson-mass}, we discuss how this model can explain the CDF anomaly.
Conversion-driven freeze-out is introduced in Sec.~\ref{sec:CDFO} and applied to our model.
We discuss our results in Sec.~\ref{sec:results} and conclude in Sec.~\ref{sec:conclusions}.

\section{Model}
\label{sec:model}

The scotogenic model~\cite{Ma:2006km} extends the SM by an $SU(2)_L\times U(1)_Y$ doublet $\eta  \sim (\vec{2},1/2) $ -- the same quantum numbers as the Higgs doublet $\Phi$ -- and (in the considered case) three right-handed singlet fermions $N_{1,2,3} \sim (\vec{1},0)$. These new particles are odd under a new $\mathbb{Z}_2$ symmetry, while the SM particles are even; this guarantees the stability of the lightest of the new particles and, thus, provides a DM candidate. 
We will focus on the scenario where DM is made up of the $N_k$ rather than scalars.

The interactions and mass terms of the right-handed fermions $N_k$ take the form
\begin{equation}
    -\mathcal{L}_N = y_{\alpha k} \Bar{L}_\alpha \tilde{\eta} N_{k} + \frac{1}{2} \sum_k m_{N_k} N_k N_k + \text{h.c.}\,,
\end{equation}
where we have already diagonalized the Majorana mass matrix of the $N$ without loss of generality.
The most general renormalizable $\mathbb{Z}_2$-symmetric scalar potential $V$ involving the SM Higgs doublet $\Phi$ and the new $\eta$ reads
\begin{align}
            V&=\mu_h^2 \Phi^\dagger \Phi + \mu_\eta^2\eta^\dagger\eta+\frac{\lambda_1}{2}(\Phi^\dagger\Phi)^2+\frac{\lambda_2}{2}(\eta^\dagger\eta)^2 \label{eq:potential}\\
            &
            +\lambda_3(\Phi^\dagger\Phi)(\eta^\dagger\eta)+\lambda_4(\Phi^\dagger\eta)(\eta^\dagger\Phi)-\left(\frac{\lambda_5}{2} (\Phi^\dagger\eta)^2+\text{h.c.}\right).
           \nonumber 
\end{align}
$\mu_h^2$ is negative to trigger electroweak symmetry breaking, but $\mu_\eta^2$ is positive and $\eta$ does not acquire a vacuum expectation value. 
Any phase of $\lambda_5$ can be absorbed into $\eta$ and eventually the lepton fields, allowing us to restrict $\lambda_5$ to non-negative values without loss of generality, similar to the inert-doublet model~\cite{Ilnicka:2015jba}.
The unbroken $\mathbb{Z}_2$ symmetry ensures that there is no mixing between the SM Higgs $h$ and the new neutral scalars in $\eta$, parametrized via
\begin{align}
\eta = \matrixx{H^+ \\ \frac{1}{\sqrt{2}} \left( H + \ii A\right)} .
\end{align}
We note that, despite the formal similarity with non-$\mathbb{Z}_2$-symmetric two-Higgs-doublet models~\cite{Branco:2011iw}, $H$ and $A$ cannot be assigned definite CP properties~\cite{Belyaev:2016lok}, because they do not have diagonal couplings to fermions.  
The masses of the scalar fields after electroweak symmetry breaking are given by~\cite{Deshpande:1977rw}
\begin{align}
    m^2_h=\lambda_1 v^2,\, \hspace{4mm} m_{H^+}^2 = \mu_\eta^2+\frac{\lambda_3}{2}v^2 ,\\
   m^2_{\Hn}=\mu_{\eta}^2+\frac{v^2}{2}\left(\lambda_3+\lambda_4- \lambda_5\right),\\
   m^2_{\An}=\mu_{\eta}^2+\frac{v^2}{2}\left(\lambda_3+\lambda_4+ \lambda_5\right).
\end{align}
Here, $v = \sqrt{-2\mu_h^2/\lambda_1}\simeq\unit[246]{GeV}$ is the vacuum expectation value of the SM Higgs doublet, with Higgs mass $m_h\simeq \unit[125]{GeV}$~\cite{Workman:2022ynf}. 
Our choice $\lambda_5\geq 0$ defines $A$ to be the heavier of the two new neutral scalars without loss of generality, with mass splitting given by $\lambda_5$: $m_A^2 - m_H^2 = \lambda_5 v^2$. For later convenience we define the linear combination of couplings
\begin{align}
\lambda_\text{L}\equiv \lambda_3+\lambda_4-\lambda_5
\end{align}
and note that some of our $\lambda$ couplings are defined differently from other articles.
The scalar couplings of Eq.~\eqref{eq:potential} have to comply with the vacuum stability conditions~\cite{Deshpande:1977rw,Kannike:2012pe}
\begin{align}
&\lambda_1 \geq 0\,,\quad
\lambda_2 \geq 0\,,\quad
\lambda_3 + \sqrt{\lambda_1\lambda_2} \geq 0\,,\\
&\lambda_3+\lambda_4-\lambda_5+\sqrt{\lambda_1\lambda_2} \geq 0\,,
\end{align}
and the perturbativity condition~\cite{Barbieri:2006dq}
\begin{align}
|2\lambda_3 (\lambda_3+\lambda_4) + \lambda_4^2 +\lambda_5^2| \lesssim 50\,.
\label{eq:pert}
\end{align}

\begin{figure}[tb]
    \centering
    \includegraphics[scale=0.5]{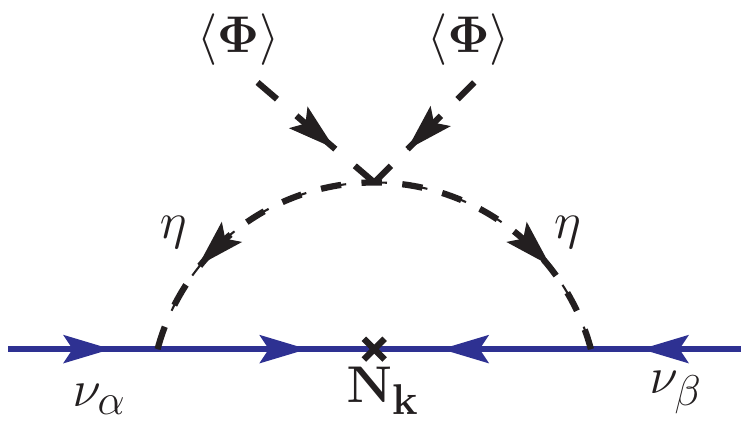}
    \vspace{-2ex}
    \caption{Radiative neutrino mass generation at one loop.
    }
    \label{fig:numass}
\end{figure}

Because of the unbroken $\mathbb{Z}_2$ symmetry of the Lagrangian, the Majorana fermions $N_k$ do not mix with the left-handed neutrinos, so the latter remain massless at tree level. However, the simultaneous presence of $y$, $\lambda_5$, and $m_N$ explicitly breaks  lepton number by two units, generating a one-loop Majorana neutrino mass matrix $\mathcal{M}_\nu$ through the diagrams shown in Fig.~\ref{fig:numass} as
\begin{align}
    &\left(\mathcal{M}_\nu\right)_{\alpha\beta}= \sum_k y_{\alpha k}^* \Lambda_{k}y^*_{\beta k}  \, , \text{ with }  \label{eq:numass}\\
    &\Lambda_k = \frac{m_{N_k}}{32\pi^2}\Bigg[\frac{m_{\Hn}^2 \log\left(\frac{m_{\Hn}^2}{m_{N_k}^2}\right)}{m_{\Hn}^2-m_{N_k}^2}  - (m_{\Hn} \leftrightarrow m_{\An})\Bigg]. 
    \label{eq:Lambda_k}
    \end{align}
(Our expression is smaller than Ma's~\cite{Ma:2006km} by a factor $1/2$, in agreement with Ref.~\cite{Merle:2015ica}.)
In addition to the loop suppression factor, neutrino masses in the scotogenic model can be suppressed in several ways, all directly linked to the restoration of lepton number and, thus, technically natural in the sense of 't~Hooft~\cite{tHooft:1979rat}: 1) small $y$, 2) small $\lambda_5$, 3) small $m_N$ or very large $m_N$.    

In the present work we are interested in the region of parameter space where the Yukawas $y_{ik}$ are small, as this enables conversion-driven freeze-out; for this we note that $|\Lambda_k|$ is maximized when the mass splitting between the scalars  $H$ and $A$ is as large as possible and $m_{N_k}$ is equal to the \emph{heavier} one, $A$.
This gives $|\Lambda_k|\simeq \sqrt{\lambda_5}\frac{v}{32\pi^2}\simeq 1\ {\rm GeV}\times \sqrt{\lambda_5}$ and, thus, a naive lower bound on the Yukawa of $|y| \gtrsim \sqrt{|\mathcal{M}_\nu/\Lambda|}\sim 5\times 10^{-6}$, assuming $\lambda_5$ is close to its perturbativity bound $\lambda_5\simeq 4$~\cite{Barbieri:2006dq}, which also requires large $\lambda_3$ and small $\lambda_4$.

A proper evaluation of the Yukawas including their flavor structure can be obtained via the Casas-Ibarra parametrization~\cite{Casas:2001sr} adapted to the scotogenic model~\cite{Toma:2013zsa}:
\begin{align}
y = U_\text{PMNS} \sqrt{\mathcal{M}_\nu^\text{diag}} R \sqrt{\Lambda^{-1}}\,,
\label{eq:CasasIbarra}
\end{align}
where $U_\text{PMNS}$ is the unitary Pontecorvo-Maki-Nakagawa-Sakata matrix and $\mathcal{M}_\nu^\text{diag}$ the diagonal matrix containing the active-neutrino masses $m_{1,2,3}$, defined through diagonalization of $\mathcal{M}_\nu$. Furthermore, $\Lambda$ is the diagonal matrix containing the $\Lambda_k$ from Eq.~\eqref{eq:Lambda_k} and $R$ is some complex orthogonal matrix. With these definitions, the above $y$ indeed solves Eq.~\eqref{eq:numass} and allows us to use neutrino masses and mixing angles as input parameters.

Neutrino oscillations have provided experimental access to the three mixing angles and the Dirac CP phase within $U_\text{PMNS}$, as well as the neutrino mass splittings~\cite{Workman:2022ynf}. The overall mass scale and the mass ordering are unknown as of yet, although there are preferences for normal ordering and a hierarchical spectrum, $m_1\ll m_{2,3}$, from global fits to oscillation data~\cite{Esteban:2020cvm} and cosmology~\cite{Planck:2018vyg,DiValentino:2021hoh}, respectively.
Assuming normal hierarchy, a vanishing lightest neutrino mass $m_1=0$, and otherwise the best-fit values for the neutrino parameters from Ref.~\cite{Esteban:2020cvm}, we find
\begin{align}
    |y| \simeq 10^{-6} \matrixx{ 0 & 2.1 & 1.4 \\ 0 & 2.4 & 7.0 \\ 0 & 2.3 & 6.1} ,
\end{align}
using a Casas-Ibarra matrix $R=\mathds{1}$, $m_H=100$\ GeV  and $m_{N_k} = m_A=266$\ GeV (i.e.~$\lambda_5=1$).
The order of magnitude matches the previous estimate but here we see that one of the $N_k$ can be arbitrarily decoupled in the limit of a vanishing lightest neutrino mass.  
For $m_1 = \unit[1]{meV}$, the first column entries are around or below $10^{-6}$; while this is the same order of magnitude as the other columns, it is just small enough to make a considerable difference for DM phenomenology.
The situation is qualitatively identical in the inverted-hierarchy case, where it is column 3 of $y$ that goes to zero with the lightest neutrino mass.

Since the Yukawas are the only portal between the SM and the $N_k$, they are crucial input parameters to study the $N$ abundance in the early Universe. For $|y_{\alpha k}|\ll 10^{-10}$, the $N_k$ would be effectively decoupled from the SM bath and, thus, nonexistent -- assuming a vanishing initial abundance after the big bang. 
The above discussion illustrates that the two measured neutrino-mass splittings enforce that at least two of the $N_k$ have couplings exceeding $10^{-6}$, which are large enough to put them in thermal equilibrium with the SM at temperatures around their masses. One of the new fermions, on the other hand, could be arbitrarily decoupled if the lightest active-neutrino mass is tiny (as hinted at by cosmology~\cite{Planck:2018vyg,DiValentino:2021hoh}).
This was already observed and employed for freeze-in fermionic DM in Refs.~\cite{Molinaro:2014lfa,Hessler:2016kwm,Baumholzer:2019twf}, for which couplings $|y|\sim 10^{-10}$ are required, considerably smaller than the Yukawas of the other two $N$.
As we will show below, keeping the couplings of the third $N$ slightly below those of the other two  -- but of similar order of magnitude -- opens up a qualitatively different DM production mechanism: conversion-driven freeze-out.

\section{\texorpdfstring{Correction to $W$-boson mass}{Wmass}}
\label{sec:W-boson-mass}

The CDF Collaboration has recently released their legacy measurement of the $W$ mass, $\unit[80.4335]{GeV}\pm \unit[9.4]{MeV}$~\cite{CDF:2022hxs}. This is not only more precise than any previous measurement, but also deviates from the SM prediction $M_W^\text{SM} = \unit[80.360]{GeV}\pm \unit[6]{MeV}$~\cite{Workman:2022ynf} by an astonishing $7\sigma$.
The deviation can be interpreted as a sign for new physics, which can impact the $W$-mass prediction through self-energy corrections encoded in the oblique parameters $S$, $T$, and $U$~\cite{Peskin:1990zt,Peskin:1991sw} via~\cite{Maksymyk:1993zm}
\begin{align}
        M_W \simeq M_W^{\rm SM} \left[1 - \frac{\alpha_{\rm EM} (S-2 c^2_W T)}{4(c_W^2-s^2_W)} +
        \frac{\alpha_{\rm EM}}{8 s^2_W} U \right] ,  
\label{MW-STU}
\end{align}
where $s_W \equiv\sin\theta_W$ and $c_W \equiv\cos\theta_W$ with the weak-mixing angle $\theta_W\simeq 29^\circ$. Since $U$ is suppressed compared to $S$ and $T$, we will neglect it in the following. The CDF measurement then prefers $S-2 c^2_W T \simeq -0.25$. Additional constraints on $S$ and $T$ come from electroweak precision observables such as $\theta_W$. We will use the results of a recent global fit including the CDF result~\cite{Asadi:2022xiy}, which finds the best-fit value $(S,T) = (0.17,0.27)$ as well as preferred $1$ and $2\sigma$ regions that we use below. Similar results have been obtained in other fits~\cite{Lu:2022bgw}.
Since the best-fit point of this global electroweak fit results in a $W$ mass within $1\sigma$ of the CDF value, we will simply refer to it as the CDF-preferred region in the following.

The oblique parameters of the scotogenic model at one loop only depend on the scalar doublet $\eta$ and are, thus, the same as those of the inert-doublet model~\cite{Deshpande:1977rw,Barbieri:2006dq, LopezHonorez:2006gr,Grimus:2008nb}. 
We shall not display them here but note that while it is not difficult to generate a large $T$ parameter through the doublet, a large positive $S$ is challenging and would require a very light $H^+$. The best-fit value for $(S,T)$ of Ref.~\cite{Asadi:2022xiy} can, thus, not be accommodated in the scotogenic model and even the $1\sigma$ region is just barely in reach. Given the $7\sigma$ discrepancy this is hardly of practical concern and still makes the scalar doublet a highly preferred extension of the SM in light of the CDF result.

\begin{figure}[tb]
    \centering
    \includegraphics[width=0.4\textwidth]{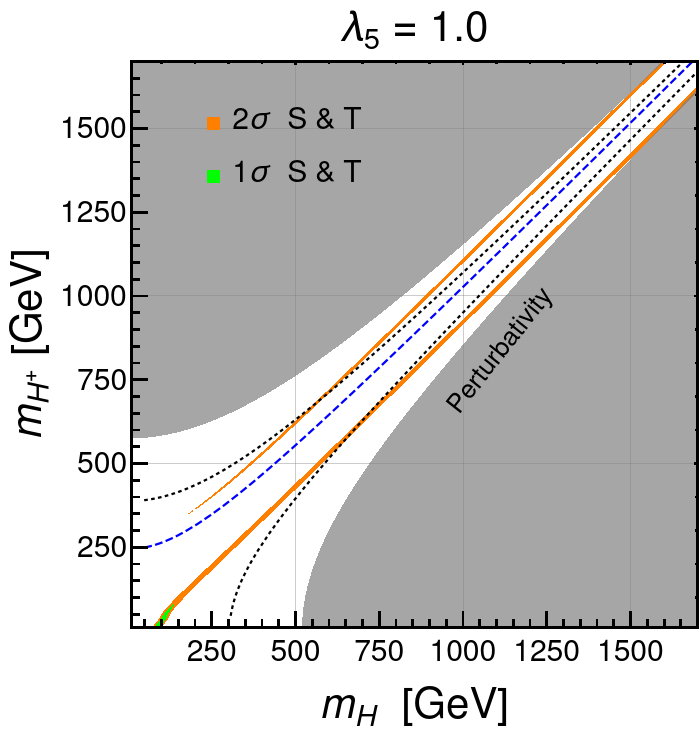}
    \caption{$1\sigma$ (green) and $2\sigma$ (orange) regions preferred by the CDF measurement of $M_W$, assuming $\lambda_5=1$.  The dashed blue line shows $m_A$. 
    The gray region is excluded by perturbativity~\cite{Barbieri:2006dq} with $|\lambda_4| \lesssim 10$, whereas the black dotted line represents the bound when $\lambda_4 \leq 4$.}
    \label{fig:scalar_hierarchies}
\end{figure}

Focusing on the small $y$ (and thus large $\lambda_5$) region of interest in this article, we show the CDF-preferred parameter space in Fig.~\ref{fig:scalar_hierarchies}. 
The $1\sigma$ region requires a rather light $H^+$ with hierarchy $m_{H^+} <m_H<m_A$, forcing $H^+$ to decay into $N_k\ell_\alpha^+$, reminiscent of supersymmetric slepton decays into neutralinos and, thus, subject to stringent limits from LEP~\cite{OPAL:2003wxm,OPAL:2003nhx},   ATLAS \cite{ATLAS:2014hep,ATLAS:2019lff,ATLAS:2019gti} and CMS~\cite{CMS:2018eqb,CMS:2018yan,CMS:2020bfa}.
At $2\sigma$ in the global fit~\cite{Asadi:2022xiy}, we have ample parameter space that survives collider constraints, including both hierarchies $m_A >  m_H > m_{H^+}$ and $ m_{H^+} > m_A >  m_H$. Even at $2\sigma$ this is a substantial improvement over the SM-only fit given the $7\sigma$ deviation of the CDF result.
Note that perturbativity and unitarity together with the CDF result ultimately put an upper bound on the masses that depends on the other $\lambda$ couplings, as can be seen in Fig.~\ref{fig:scalar_hierarchies}.

\section{Conversion-driven freeze-out }
\label{sec:CDFO}

Conversion-driven freeze-out~\cite{Garny:2017rxs} (or co-scattering~\cite{DAgnolo:2017dbv}) can be realized in regions of parameter space where the preservation of chemical equilibrium through efficient conversions in the $\mathbb{Z}_2$-odd sector leads to an underabundance of DM\@. In this region, sufficiently small conversion rates between DM (here taken to be $N_1$) and its coannihilating partners can initiate the chemical decoupling of DM and hinder its efficient dilution due to coannihilation effects. As the chemical decoupling of and within the $\mathbb{Z}_2$-odd sector may overlap in time, in general, abundances of all $\mathbb{Z}_2$-odd particles need to be tracked individually; \ie, a coupled set of Boltzmann equations need to be solved:
\bea
\label{eq:BMEgen}
  \frac{\dd Y_i}{\dd x} &= &\frac{1}{ 3 \mathcal{H}}\frac{\dd s}{\dd x}
  \left[\,\sum_j\avb{\sigma_{ij\to\mathrm{SM}} v}\left(Y_i Y_j-Y_i^{\eq}Y_j^{\eq}\right)\right.\nonumber
  \\
   & &\qquad\qquad \left.+\sum_{j\neq i} \frac{\Gamma_{\text{conv}}^{i\to j} }{s}\left(Y_i-Y_j\frac{Y_i^{\eq}}{Y_j^{\eq}}\right)\right],
\ena
where $i,j\in \{\Hn,\An,\Hpm,N_1,N_2,N_3\}$. Here, $Y_i$ denotes the comoving number density of species $i$, $x=m_{N_1}/T$ is the temperature parameter (with $T$ being the temperature of the SM thermal bath), $\mathcal{H}$ the Hubble rate, and $s$ the entropy density. Furthermore,   $\avb{\sigma_{ij\to\mathrm{SM}} v}$ denotes the thermally averaged annihilation cross section times M{\o}ller velocity, $\Gamma_{\text{conv}}^{i\to j}$ the sum of the conversion rates for scatterings and (inverse) decays, and $Y_i^\eq$ the equilibrium density; see \eg~\cite{Garny:2017rxs,Edsjo:1997bg} for explicit expressions for these quantities. 
We have neglected annihilations within the $\mathbb{Z}_2$-odd sector, as they are of higher order in the involved couplings and are negligible compared to the conversions in the region of parameter space where chemical equilibrium within the $\mathbb{Z}_2$-odd sector is questionable.

In the scotogenic model, we can simplify Eq.~\eqref{eq:BMEgen}. The sizable couplings within the scalar doublet $\eta$ due to its gauge interactions render the conversions among them efficient, and we can assume
\beq
\frac{Y_i}{Y_i^\eq}=\frac{Y_j}{Y_j^\eq}
\eeq    
for $i,j\in \{\Hn,\An,\Hpm\}$, \ie, apply the well-known coannihilation approximation~\cite{Edsjo:1997bg}. We are left with the set of $1+3$ Boltzmann equations, one  for $\YID=\sum_i Y_i, \; i\in \{\Hn,\An,\Hpm\}$ and one for each $Y_{N_k}$:
\begin{align}
\label{eq:BMEID}
\begin{split}
  \frac{\dd \YID}{\dd x} &= \frac{1}{ 3 \mathcal{H}}\frac{\dd s}{\dd x}
  \left[\,\avb{\sigma v}_\mathrm{eff}\left(\YID^2- (\YID^{\eq})^2\right)\right.\\
   & \quad + \left.\sum_k\frac{\Gamma_{\text{conv,\,eff}}^{\eta\to {N_k}} }{s}\left(\YID-Y_{N_k}\frac{\YID^{\eq}}{Y_{N_k}^{\eq}}\right)\right],
   \end{split}\\
   \label{eq:BMEN}
  \frac{\dd Y_{N_k}}{\dd x} &= - \frac{1}{ 3 s \mathcal{H}}\frac{\dd s}{\dd x}\,\Gamma_{\text{conv,\,eff}}^{\eta\to {N_k}} \left(\YID-Y_{N_k}\frac{\YID^{\eq}}{Y_{N_k}^{\eq}}\right),
\end{align}
where we employed the commonly used effective cross section
\beq
\avb{\sigma v}_\mathrm{eff}\equiv\sum_{i,j\in \{\Hn,\An,\Hpm\}} \avb{\sigma_{ij\to\mathrm{SM}} v}\frac{Y_i^\eq Y_j^\eq}{(Y_\eta^\eq)^2}\,.
\label{eq:effxs}
\eeq
As we cannot assume efficient conversions between the doublet states and $N_k$, the respective terms do not cancel out when summing up the Boltzmann equations for the scalars.
Similar to the effective cross section, they can, however, be expressed as an effective conversion rate: 
\beq
\Gamma_{\text{conv,\,eff}}^{\eta\to N_k}\equiv \sum_{i\in \{\Hn,\An,\Hpm\}} \Gamma_{\text{conv}}^{i\to N_k}\frac{Y_i^\eq }{Y_\eta^\eq}\,.
\label{eq:effGammaConv}
\eeq
Note that the annihilation rate of $N_k$ can safely be neglected for the small Yukawa couplings compatible with conversion-driven freeze-out. 

The decay width of the charged scalar $H^+$ into a charged lepton $\ell$ and a Majorana fermion $N$ reads as  
\begin{align}
\begin{split}
    \Gamma^{H^+ \to \ell_i N_k} &= \frac{|y_{i k}|^2\ m_{H^+}}{16 \pi} \left( 1 - \frac{m_{N_k}^2}{m_{H^+}^2} - \frac{m_{\ell_i}^2}{m_{H^+}^2} \right)\\
    &\quad\times 
    F\left[ \frac{m_{N_k}^2}{m_{H^+}^2}, \frac{m_{\ell_i}^2}{m_{H^+}^2} \right]^{1/2} 
\end{split}
\label{eq:decayrate}
\end{align}    
with the function 
\begin{equation}
    F[a,b] \equiv 1 + a^2 + b^2 - 2 a b - 2 a -2 b\,.
\end{equation}
The decay widths for the neutral scalars $\{H,A\}$ into $\nu_i N_k$ are obtained by simply replacing the respective masses in the above expression, $m_{H^+} \rightarrow m_{H,A}$ and $m_\ell \rightarrow m_\nu$.  

For the numerical solution of the Boltzmann equations \eqref{eq:BMEID} and \eqref{eq:BMEN}, 
we compute $\avb{\sigma v}_\mathrm{eff}$ with micrOMEGAs~\cite{Belanger:2014vza}, employing  CalcHep~\cite{Pukhov:2004ca} and the implementation of the inert-doublet model included in the program package~\cite{LopezHonorez:2010eeh}.
For the computation of the conversion rate, we take into account only the contribution of (inverse) decays following Eq.~\eqref{eq:decayrate}.

In general, conversions via scattering can be important, in particular for small $x$, \ie~at an early state of the freeze-out process~\cite{Garny:2017rxs}. However, as we will show in Sec.~\ref{sec:results}, in the region of interest, the mass splitting between $N_k$ and the lightest doublet state is small compared to the mass splitting within the doublet. Accordingly, the considerably stronger phase-space suppression of the decay of the lightest inert state typically renders the respective rate several orders of magnitude smaller than the one for the heavier states. Hence, according to Eq.~\eqref{eq:effGammaConv}, the contribution to $\Gamma_{\text{conv,\,eff}}^{\eta\to N_k}$ from decays of heavier states dominates over the one of the lightest state at small $x$, \ie~for $x$ compatible with or smaller than the inverse of the relative mass splitting within the doublet. As a result, scatterings are significantly less relevant at an early state of the freeze-out process where they could potentially dominate. In fact, they can safely be neglected for the bulk of the parameter space of interest as we have checked by explicitly taking into account their leading contributions for several benchmark points.\footnote{We have performed two independent checks. On the one hand, we have computed the scattering cross sections with MadGraph5\_aMC@NLO/MadDM~\cite{Alwall:2014hca,Ambrogi:2018jqj} and solved the Boltzmann equations \eqref{eq:BMEID} and \eqref{eq:BMEN} accordingly. On the other hand, the most recent version of micrOMEGAs~\cite{Alguero:2022inz} directly allows for the computation of the relic density in the case of conversion-driven freeze-out. Note, however, that the latter has provided no or obviously incorrect results (due to double counting of decays and scatterings in the presence of resonances) in some cases. For both checks, we have used SARAH~\cite{Staub:2008uz,Staub:2009bi} for the implementation of the scotogenic model in the appropriate format. }
Scatterings can, however, become important for very small mass splittings $\Delta m$, below $\sim 1\,$GeV for $m_\Hpm<m_H,m_A$ and considerably smaller values for $m_H<m_\Hpm,m_A$. 

Further corrections that can affect our result quantitatively come from non-perturbative effects such as Sommerfeld enhancement and bound state effects. In particular, the latter can significantly enlarge the viable region of conversion-driven freeze-out for a strongly interacting coannihilator as recently found in~\cite{Garny:2021qsr}. We expect a qualitatively similar but smaller effect here. Their study is, however, beyond the scope of this work.

In writing down Eq.~\eqref{eq:BMEgen}, we have also implicitly assumed that the DM momentum distribution is sufficiently close to the thermal distribution with temperature $T$. While kinetic equilibrium is not guaranteed to be maintained through efficient elastic scatterings, this assumption is, nevertheless, expected to hold. Because of the small mass splitting between DM and the coannihilating particles, the thermal distribution of the latter is inherited by DM, as discussed in the appendix of Ref.~\cite{Garny:2017rxs}. In that study, the unintegrated Boltzmann equations have been solved for conversion-driven freeze-out with similar mass scales, implying a theoretical error of $\lesssim 10\%$ induced by the above assumption.

\section{Results and Discussion}
\label{sec:results}

\begin{figure*}
  \centering
\includegraphics[width=0.45\textwidth,trim={0 0.1cm 0 0},clip]{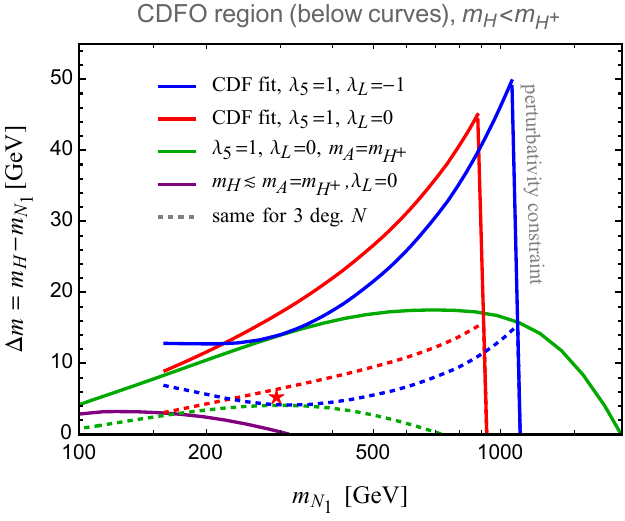}
\hspace{3ex}
\includegraphics[width=0.472\textwidth,trim={0 0 0 0.08cm},clip]{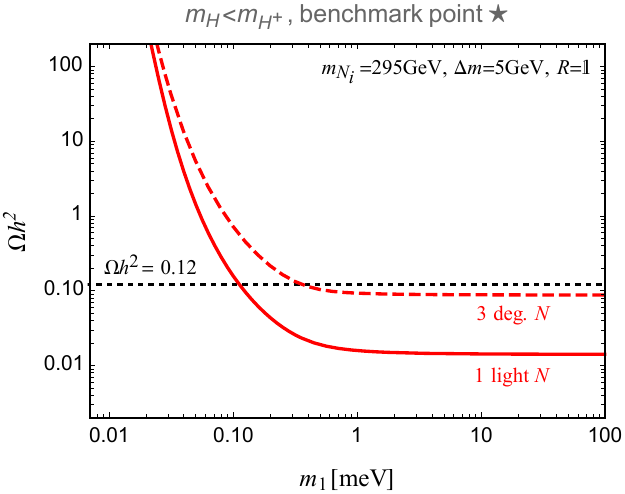}
  \caption{
 Left: Boundary of the conversion-driven freeze-out region for the case $m_{N_1}<m_H < m_A,m_{H^+}$ for four different slices in the parameter space as a function of DM mass $m_{N_1}$ and mass splitting $\Delta m = m_H-m_{N_1}$.\\
Right: Relic density as a function of the lightest active neutrino mass $m_1$ for the benchmark point shown as a red star in the left panel for the choice of one light $N$ only, $m_{N_{2,3}}=m_A$ (solid curve), and three mass-degenerate $N$, $m_{N_{2,3}}=m_{N_1}$ (dashed curve).
 }
  \label{fig:plotsHnL}
\end{figure*}

As explained in Sec.~\ref{sec:CDFO}, the viable region of conversion-driven freeze-out requires underabundant DM if chemical equilibrium was maintained within the $\mathbb{Z}_2$-odd sector. We can, hence, compute the boundary of that region by the requirement to saturate the relic density under the assumption of chemical equilibrium while neglecting annihilation channels with $N_k$ in the initial states. The result for $m_{N_1} < m_H<m_A<m_\Hpm$ is shown in the left panel in Fig.~\ref{fig:plotsHnL}, where we consider four setups regarding the parameters of the scalar sector:
\begin{enumerate}
    
    \item[(i)] CDF preferred, Higgs-philic: 
    
    $\lambda_5=1$, $m_H,m_\Hpm$ according to the $2\sigma$-band in Fig.~\ref{fig:scalar_hierarchies}, $\lambda_\text{L}=-1$ (blue line)
    
    \item[(ii)] CDF preferred, Higgs-phobic: 
    
    $\lambda_5=1$, $m_H,m_\Hpm$ according to the $2\sigma$-band in Fig.~\ref{fig:scalar_hierarchies}, $\lambda_\text{L}=0$ (red line)
    
    \item[(iii)] partly mass-degenerate case: 
    
    $m_A=m_\Hpm$,  $m_H-m_A$ sizable, due to $\lambda_5=1$, $\lambda_\text{L}=0$ (green line)
    
    \item[(iv)] nearly mass-degenerate case: 
    
    $m_A=m_\Hpm=m_H+5\,$GeV, $\lambda_\text{L}=0$ (purple line)
\end{enumerate}
In scenario (i),  the lightest doublet scalar, $H$, has a large coupling $\lambda_L=-1$ to the SM Higgs $h$, which we denote as Higgs-philic; in scenario (ii), the coupling of $H$ to $h$ is turned off, rendering it Higgs-phobic.
The solid curves denote the case where only $N_1$ is lighter than $H$ and, thus, the DM\@, whereas $N_{2,3}$ are chosen sufficiently heavy to not affect the relic density.\footnote{For definiteness, we choose $N_{2,3}=m_A$; see also discussion in Sec.~\ref{sec:model}. However, for the considered magnitude of Yukawa couplings, $N_{2,3}$ are basically decoupled once their mass is about 10\% larger than the lightest scalar.} The short dashed curves denote the case of mass-degenerate fermions, so all three $N_k$ form DM [not present for scenario (iv)].
Note that the boundary for any other configuration, for instance, $m_{N_1}=m_{N_2}$ and  $m_{N_3}$ heavy, or $m_{N_1}<m_{N_{2,3}} \lesssim m_H$, lies between these two limiting cases. 
The region \emph{below} the curves in the left panel of Fig.~\ref{fig:plotsHnL} would lead to underabundant DM if conversion processes would remain efficient, $\Gamma_\text{conv}\gg \mathcal{H}$.

\begin{figure*}
  \centering
\includegraphics[width=0.47\textwidth,trim={0 0.1cm 0 0},clip]{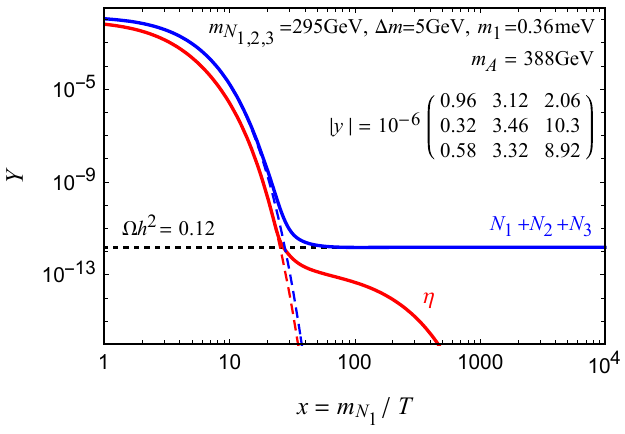}
\hspace{3ex}
\includegraphics[width=0.47\textwidth,trim={0 0.1cm 0 0},clip]{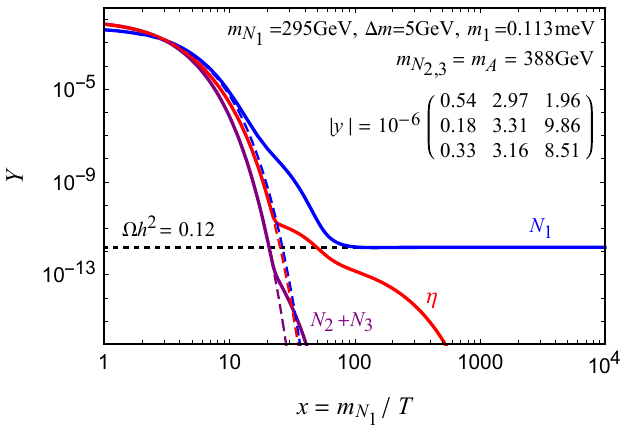}
  \caption{
 Evolution of the DM abundance $Y_N$ (blue line) and the  doublet abundance $\YID$ (red line) for the benchmark point considered in the right panel of Fig.~\ref{fig:plotsHnL}, taking $m_1$ to match the measured relic density, $\Omega h^2 = 0.12$. The left panel represents the case of three mass-degenerate $N$. For the right panel, only $N_1$ is assumed to be light, while $m_{N_2}=m_{N_3}=m_A$. The purple curves in the right panel show the abundance of the heavier $N$. The parameters of the doublet are $m_\Hn=300\,$GeV, $m_\An\simeq388\,$GeV, $m_\Hpm\simeq442\,$GeV, and $\lambda_\text{L}=0$. The dashed lines show the equilibrium abundances, solid lines show the solution of the coupled Boltzmann equations. Only conversions via (inverse) decays are taken into account here, see the text for details.
 }
  \label{fig:abundances}
\end{figure*}

Conversion-driven freeze-out evidently requires a moderately small mass splitting between DM and the next-to-lightest  $\mathbb{Z}_2$-odd scalar, $H$ in Fig.~\ref{fig:plotsHnL} (left).
We make the interesting observation that sizable mass splittings within the doublet (due to $\lambda_5=1$ and -- to a larger extent -- due to additionally fitting the CDF measurement) significantly enlarge the viable region for conversion-driven freeze-out. This is due to 
destructive interferences among the various diagrams of annihilation processes of the scalars that suppress their cross section [and overcompensate the competing Boltzmann suppression of heavier states in Eq.~\eqref{eq:effxs}] see \eg~discussion in Refs.~\cite{Lundstrom:2008ai,
LopezHonorez:2010tb,Eiteneuer:2017hoh}. The relatively large mass splitting between $H$ and $\Hpm$ for setups (i) and (ii) preferred by the CDF measurement require a very large $\lambda_4$ for large masses, challenging the perturbativity. The steep drop of the corresponding curves toward the right in Fig.~\ref{fig:plotsHnL} (left)  stems from imposing Eq.~\eqref{eq:pert}. Note that vacuum stability does not impose further constraints as long as we allow for $\lambda_2\gtrsim 2$.

In the right panel of Fig.~\ref{fig:plotsHnL}, we consider the benchmark point
with $m_H=300\,$GeV and $\Delta m=5\,$GeV (denoted by an asterisk in the left panel of the same figure) that belongs to the setup (ii). We use the mass of the lightest active neutrino, $m_1$, to parametrize the Yukawa couplings following Eq.~\eqref{eq:CasasIbarra} and the prescription below that formula; see Sec.~\ref{sec:model} for further details. We assume normal hierarchy. Solving the set of Boltzmann equations \eqref{eq:BMEID} and \eqref{eq:BMEN}, we compute $\Omega h^2$ as a function of $m_1$ for the cases light $N_1$ and mass degenerate $N_{1,2,3}$. The respective points that yield the observed value $\Omega h^2=0.12$~\cite{Planck:2018vyg} are considered in Fig.~\ref{fig:abundances} where we show the evolution of the comoving number densities during freeze-out. We also display the corresponding Yukawa matrices. The first column governs the interaction of $N_1$ and is somewhat smaller than the one for $N_{2,3}$. As we assumed $R=\mathbb{1}$, here, we have a direct relation between these Yukawas and $m_1$. Sizable off-diagonal elements in $R$, however, would lead to potentially larger entries in the first column and could easily lead to chemical equilibrium and, hence, $\Omega h^2<0.12$ for the chosen point inside the conversion-driven freeze-out region. Therefore, the requirement of saturating the relic density constraint for a given point inside the conversion-driven freeze-out region (\ie~for a point below the curves in the left panel of Fig.~\ref{fig:plotsHnL}) sets an upper limit on $m_1$, since $R\neq\mathbb{1}$ can only lower the relic density.
Typical values for $m_1$ lie between $\sim1\,$meV (for the smallest considered masses and three degenerate sterile neutrinos) and $\sim0.03\,$meV (for a DM mass around $1\,$TeV and the case of one light $N_1$ only).

The Yukawa matrices displayed in Fig.~\ref{fig:abundances} contain values that span roughly 1.5 orders of magnitude. For smaller (larger) $m_H$ and otherwise compatible parameters, the spread among the entries of $|y|$ become smaller (larger) requiring $\Omega h^2=0.12$.
Conversion-driven freeze-out, thus, requires small Yukawas -- similar to that of the electron -- but very little hierarchy in the matrix.
This conclusion also holds for inverted neutrino hierarchy.

\begin{figure}
  \centering
\includegraphics[width=0.45\textwidth,trim={0 0 0 0},clip]{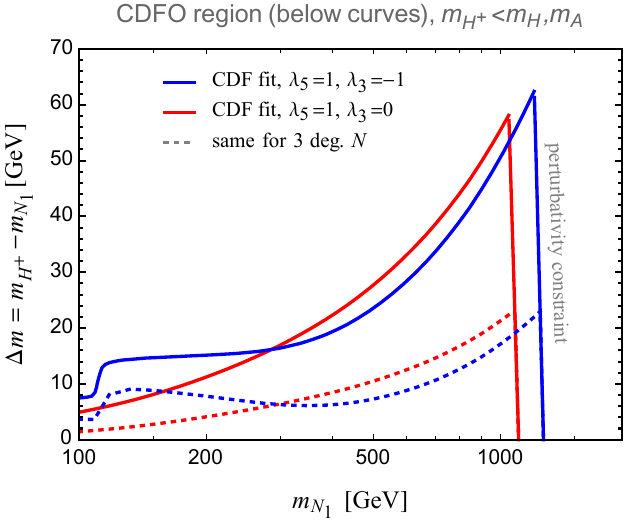}\\[1.5ex]
\includegraphics[width=0.466\textwidth,trim={0.05cm 0.08cm 0.05cm  0},clip]{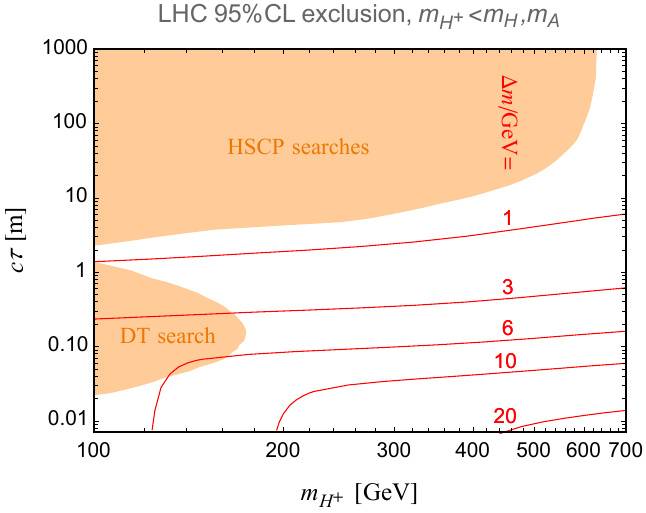}
  \caption{
 Results for the scenario $m_{H^+}<m_H,m_A$. \\
 Top: Boundary of the conversion-driven freeze-out region for one light $N$ (solid lines) and degenerate $N$ (short dashed). \\
 Bottom: LHC constraints from long-lived particle searches for disappearing tracks (DT search) and heavy stable charged particles (HSCP searches) for the case $\lambda_3=0$, \ie~scenario~(v). The red lines denote contours of constant $\Delta m$ for one light $N$. We choose $R=\mathbb{1}$, and solve for $m_1$ such as to match $\Omega h^2=0.12$.
}
  \label{fig:plotsHpL}
\end{figure}

So far, we have focused on the mass hierarchy $m_{N_1} < m_H<m_A<m_\Hpm$, let us now briefly consider $m_{N_1} <m_\Hpm<m_H<m_A$.
The boundaries for conversion-driven freeze-out for this case are shown in the top panel of Fig.~\ref{fig:plotsHpL}. They are qualitatively similar to the ones shown in Fig.~\ref{fig:plotsHnL}. To reduce clutter, we therefore display only the curves for the two setups preferred by the CDF measurements:
\begin{enumerate}
    \item[(v)] CDF preferred, Higgs-phobic: 
    
    $\lambda_5=1$, $m_H,m_\Hpm$ according to the $2\sigma$-band in Fig.~\ref{fig:scalar_hierarchies}, $\lambda_\text{3}=0$ (red line)
    
    \item[(vi)] CDF preferred, Higgs-philic: 
    
    $\lambda_5=1$, $m_H,m_\Hpm$ according to the $2\sigma$-band in Fig.~\ref{fig:scalar_hierarchies}, $\lambda_\text{3}=-1$ (blue line)
\end{enumerate}
Note that the $2\sigma$-band of the CDF anomaly (Fig.~\ref{fig:scalar_hierarchies}) extends to smaller masses for this hierarchy, allowing us to explore masses down to 100\,GeV. In particular, we observe a bump in $\Delta m$ around $m_\Hpm\sim m_h$ due to the threshold of the annihilation channel $H^- \Hpm\to h h$. As $\lambda_3$ governs the Higgs-portal coupling of $\Hpm$, this bump occurs only for the blue curve.

We stress that the viable realization of conversion-driven freeze-out with Yukawa couplings of the order of around $10^{-6}$ is not restricted to the region preferred by the CDF measurement, although we focus on these regions for the choice of our concrete examples. In fact, qualitatively similar results can be found for the partly mass-degenerate doublet case (iii). A further region worth mentioning is the Higgs resonant region where $m_H\simeq m_h/2$. For sizable $\lambda_\text{L}$, resonant $H$ pair annihilation provides a large cross section allowing for  conversion-driven freeze-out solutions for a sizable range of $\Delta m$.

\subsection{LHC constraints}

To derive constraints on the scotogenic model from searches for new physics at the LHC, we use SModelS~2.2.1~\cite{Alguero:2021dig}, employing the interface of micrOMEGAs~\cite{Barducci:2016pcb} for the computation of cross sections and decay tables. Since the $N_k$ are gauge singlets and have small Yukawas, their production rates are heavily suppressed compared to the doublet scalars. For the case $m_H < m_A,m_{H^+}$, the signature of the model is missing transverse energy since the heavier scalars promptly decay into the neutral $H$. As the lightest doublet scalar $H$ predominantly decays into a pair of neutral particles (an active neutrino and $N$) its decay length does not affect its signature. 
Such missing-energy signatures are typical for a wide variety of DM models.
Interestingly, this part of the parameter space of interest is not challenged by any search implemented in SModelS. The most relevant searches are Refs.~\cite{ATLAS:2021moa,ATLAS:2019lff,ATLAS:2019wgx,CMS:2021edw,CMS:2017moi}.

In contrast, the scenario $m_{H^+} < m_A,m_H$ can lead to striking signatures of long-lived particles at the LHC\@, as $H^+$ can decay only through the suppressed Yukawa interactions. With the small Yukawas and mass splittings required for conversion-driven freeze-out, the lifetime of $H^+$ can become very long. For decay lengths larger than the size of the detector, $H^+$ is likely to traverse the entire detector leading to the signature of highly ionizing tracks and anomalous time-of-flight measurements, commonly denoted as heavy stable charged particles (HSCPs). For decay lengths compatible with the size of the inner detector, disappearing track or displaced lepton searches are most sensitive. Figure~\ref{fig:plotsHpL} (bottom) shows the respective exclusion at 95\% CL, obtained from SModelS\@. The two isolated exclusion regions above and below  $\sim 1\,$m stem from HSCP~\cite{ATLAS:2019gqq, CMS:2015lsu, CMS:2013czn} and  disappearing track~\cite{ATLAS:2017oal} searches, respectively. In addition, we display five contours of constant $\Delta m$ in the conversion-driven freeze-out region solving for $\Omega h^2=0.12$. The region close to its boundary corresponds to the smallest lifetimes shown. On the other hand, HSPC searches are sensitive only for $\Delta m$ well below a GeV. As can be seen, intermediate values for $\Delta m$ provide an observable signal in disappearing track searches, already excluding a small portion of the parameter. 

Toward larger $\Delta m$ -- and thus smaller lifetimes -- disappearing-track searches are less sensitive, requiring a hit in a minimal number of layers in the inner detector. Displaced lepton searches are more promising to tackle this region of parameter space. However, the ATLAS displaced lepton search, Ref.~\cite{ATLAS:2020wjh}, implemented in SModelS, does not provide sensitivity, being interpreted for a large mass splitting between the mother particle and DM and, hence, for relatively hard leptons. In our considered scenario, the displaced lepton from $H^-\to N \ell$ is rather soft due to the small mass splitting $\Delta m$.

While very large decay lengths (relevant for very small $\Delta m$ only) are well constrained by HSCP searches, there is a considerable gap between the disappearing track and HSCP search, again, leaving the entire mass range unconstrained. 
Dedicated searches for the particular scenario taking into account the highly ionizing nature of $H^+$ as well as the displaced decay into a lepton and missing energy are expected to greatly enhance the sensitivity.

\subsection{Other constraints and predictions}

Conversion-driven freeze-out in the scotogenic model requires fermionic DM, so DM interacts with the SM exclusively through the Yukawa couplings $y$. The couplings $|y_{\alpha k}|$ of the $N_k$ that forms DM have to be of the order of $10^{-7}$--$10^{-6}$ for conversion to work. Couplings of this magnitude and masses in the \unit[100]{GeV} to TeV region automatically suppress any and all DM direct and indirect detection signatures. 

Since $|y_{\alpha k}|$ for $N_k$ below $10^{-6}$ require a lightest neutrino mass around or below meV, this is a prediction of our scenario. This upper limit on the lightest Majorana neutrino mass also implies a lower limit on the rate for neutrinoless double-beta decay, seeing as that rate can vanish only for $m_1\in \unit[(1,10)]{meV}$ under normal ordering~\cite{Rodejohann:2011mu}.

The heavier $N_k$ could still have an impact, for example on lepton flavor violation, if their Yukawa couplings were large. This is, in principle, allowed in the conversion-driven region but would imply a large hierarchy in the $y$ matrix columns, only to be constrained by lepton flavor violation. The more natural region has all $y$ entries of the order of $10^{-6}$, which then suppresses lepton flavor violation and all other signatures involving the neutral fermions.

In a special region of parameter space the $N_k$ are almost degenerate, leading to one or two long-lived fermions that decay via $N\to \ell^+\ell^-N_\text{DM}$, $N\to \nu\nu N_\text{DM}$, or $N\to \gamma N_\text{DM}$. Such scenarios could be constrained by big bang nucleosynthesis if the lifetimes are long enough, but shall not be discussed further here.

\vspace{-2ex}
\section{Conclusions}
\label{sec:conclusions}

The scotogenic model is a well-motivated extension of the SM that simultaneously addresses DM and the smallness of neutrino masses. The included scalar doublet also allows for an amelioration of the CDF anomaly by contributing positively to the $W$ mass. DM production via freeze-out and freeze-in has been discussed in the literature at length; here we showed for the first time that the scotogenic model also allows for DM production via conversion-driven freeze-out. This requires fermionic DM with Yukawa couplings in between the typical freeze-out and freeze-in values. In particular, the necessary  Yukawa matrix is neither tiny nor hierarchical, and the same is true for the parameters of the scalar potential. Conversion-driven freeze-out is, hence, quite natural.

The lightest active-neutrino mass plays a crucial role and is typically between 0.03 and \unit[1]{meV}.
Such a hierarchical active-neutrino spectrum is becoming increasingly more probable to satisfy cosmological limits on the sum of neutrino masses.
Parts of the parameter space are furthermore testable at colliders, notably through displaced-vertex decays of the charged scalar.

\section*{Acknowledgements}

This work was supported in part by the National Science Foundation under Grant No.~PHY-2210428.
J.~Heisig acknowledges support by the Alexander von Humboldt foundation via the Feodor Lynen Research Fellowship for Experienced Researchers. 

\bibliographystyle{utcaps_mod}
\bibliography{bib}

\providecommand{\href}[2]{#2}\begingroup\raggedright\begin{thebibliography}{10}

\bibitem{Ma:2006km}
E.~Ma, ``{Verifiable radiative seesaw mechanism of neutrino mass and dark
  matter},'' \href{http://dx.doi.org/10.1103/PhysRevD.73.077301}{{\em Phys.
  Rev. D} {\bfseries 73} (2006) 077301},
  \href{http://arxiv.org/abs/hep-ph/0601225}{[{\ttfamily hep-ph/0601225}]}.

\bibitem{Deshpande:1977rw}
N.~G. Deshpande and E.~Ma, ``{Pattern of Symmetry Breaking with Two Higgs
  Doublets},'' \href{http://dx.doi.org/10.1103/PhysRevD.18.2574}{{\em Phys.
  Rev. D} {\bfseries 18} (1978) 2574}.

\bibitem{Barbieri:2006dq}
R.~Barbieri, L.~J. Hall, and V.~S. Rychkov, ``{Improved naturalness with a
  heavy Higgs: An Alternative road to LHC physics},''
  \href{http://dx.doi.org/10.1103/PhysRevD.74.015007}{{\em Phys. Rev. D}
  {\bfseries 74} (2006) 015007},
  \href{http://arxiv.org/abs/hep-ph/0603188}{[{\ttfamily hep-ph/0603188}]}.

\bibitem{LopezHonorez:2006gr}
L.~Lopez~Honorez, E.~Nezri, J.~F. Oliver, and M.~H.~G. Tytgat, ``{The Inert
  Doublet Model: An Archetype for Dark Matter},''
  \href{http://dx.doi.org/10.1088/1475-7516/2007/02/028}{{\em JCAP} {\bfseries
  02} (2007) 028}, \href{http://arxiv.org/abs/hep-ph/0612275}{[{\ttfamily
  hep-ph/0612275}]}.

\bibitem{Borah:2020wut}
D.~Borah, A.~Dasgupta, K.~Fujikura, S.~K. Kang, and D.~Mahanta, ``{Observable
  Gravitational Waves in Minimal Scotogenic Model},''
  \href{http://dx.doi.org/10.1088/1475-7516/2020/08/046}{{\em JCAP} {\bfseries
  08} (2020) 046}, \href{http://arxiv.org/abs/2003.02276}{[{\ttfamily
  2003.02276}]}.

\bibitem{deBoer:2021pon}
T.~de~Boer, R.~Busse, A.~Kappes, M.~Klasen, and S.~Zeinstra, ``{Indirect
  detection constraints on the scotogenic dark matter model},''
  \href{http://dx.doi.org/10.1088/1475-7516/2021/08/038}{{\em JCAP} {\bfseries
  08} (2021) 038}, \href{http://arxiv.org/abs/2105.04899}{[{\ttfamily
  2105.04899}]}.

\bibitem{Liu:2022byu}
J.~Liu, Z.-L. Han, Y.~Jin, and H.~Li, ``{Unraveling the Scotogenic Model at
  Muon Collider},'' \href{http://arxiv.org/abs/2207.07382}{[{\ttfamily
  2207.07382}]}.

\bibitem{McDonald:2001vt}
J.~McDonald, ``{Thermally generated gauge singlet scalars as selfinteracting
  dark matter},'' \href{http://dx.doi.org/10.1103/PhysRevLett.88.091304}{{\em
  Phys. Rev. Lett.} {\bfseries 88} (2002) 091304},
  \href{http://arxiv.org/abs/hep-ph/0106249}{[{\ttfamily hep-ph/0106249}]}.

\bibitem{Asaka:2005cn}
T.~Asaka, K.~Ishiwata, and T.~Moroi, ``{Right-handed sneutrino as cold dark
  matter},'' \href{http://dx.doi.org/10.1103/PhysRevD.73.051301}{{\em Phys.
  Rev. D} {\bfseries 73} (2006) 051301},
  \href{http://arxiv.org/abs/hep-ph/0512118}{[{\ttfamily hep-ph/0512118}]}.

\bibitem{Hall:2009bx}
L.~J. Hall, K.~Jedamzik, J.~March-Russell, and S.~M. West, ``{Freeze-In
  Production of FIMP Dark Matter},''
  \href{http://dx.doi.org/10.1007/JHEP03(2010)080}{{\em JHEP} {\bfseries 03}
  (2010) 080}, \href{http://arxiv.org/abs/0911.1120}{[{\ttfamily 0911.1120}]}.

\bibitem{Molinaro:2014lfa}
E.~Molinaro, C.~E. Yaguna, and O.~Zapata, ``{FIMP realization of the scotogenic
  model},'' \href{http://dx.doi.org/10.1088/1475-7516/2014/07/015}{{\em JCAP}
  {\bfseries 07} (2014) 015}, \href{http://arxiv.org/abs/1405.1259}{[{\ttfamily
  1405.1259}]}.

\bibitem{Hessler:2016kwm}
A.~G. Hessler, A.~Ibarra, E.~Molinaro, and S.~Vogl, ``{Probing the scotogenic
  FIMP at the LHC},'' \href{http://dx.doi.org/10.1007/JHEP01(2017)100}{{\em
  JHEP} {\bfseries 01} (2017) 100},
  \href{http://arxiv.org/abs/1611.09540}{[{\ttfamily 1611.09540}]}.

\bibitem{Baumholzer:2019twf}
S.~Baumholzer, V.~Brdar, P.~Schwaller, and A.~Segner, ``{Shining Light on the
  Scotogenic Model: Interplay of Colliders and Cosmology},''
  \href{http://dx.doi.org/10.1007/JHEP09(2020)136}{{\em JHEP} {\bfseries 09}
  (2020) 136}, \href{http://arxiv.org/abs/1912.08215}{[{\ttfamily
  1912.08215}]}.

\bibitem{Garny:2017rxs}
M.~Garny, J.~Heisig, B.~L\"ulf, and S.~Vogl, ``{Coannihilation without chemical
  equilibrium},'' \href{http://dx.doi.org/10.1103/PhysRevD.96.103521}{{\em
  Phys. Rev. D} {\bfseries 96} (2017) 103521},
  \href{http://arxiv.org/abs/1705.09292}{[{\ttfamily 1705.09292}]}.

\bibitem{DAgnolo:2017dbv}
R.~T. D'Agnolo, D.~Pappadopulo, and J.~T. Ruderman, ``{Fourth Exception in the
  Calculation of Relic Abundances},''
  \href{http://dx.doi.org/10.1103/PhysRevLett.119.061102}{{\em Phys. Rev.
  Lett.} {\bfseries 119} (2017) 061102},
\href{http://arxiv.org/abs/1705.08450}{[{\ttfamily 1705.08450}]}.

\bibitem{CDF:2022hxs}
{\bfseries CDF} Collaboration, T.~Aaltonen {\em et~al.}, ``{High-precision
  measurement of the W boson mass with the CDF II detector},''
  \href{http://dx.doi.org/10.1126/science.abk1781}{{\em Science} {\bfseries
  376} no.~6589, (2022) 170--176}.

\bibitem{Batra:2022pej}
A.~Batra, S.~K. A, S.~Mandal, H.~Prajapati, and R.~Srivastava, ``{CDF-II $W$
  Boson Mass Anomaly in the Canonical Scotogenic Neutrino-Dark Matter Model},''
  \href{http://arxiv.org/abs/2204.11945}{[{\ttfamily 2204.11945}]}.

\bibitem{Fan:2022dck}
Y.-Z. Fan, T.-P. Tang, Y.-L.~S. Tsai, and L.~Wu, ``{Inert Higgs Dark Matter for
  CDF II W-Boson Mass and Detection Prospects},''
  \href{http://dx.doi.org/10.1103/PhysRevLett.129.091802}{{\em Phys. Rev.
  Lett.} {\bfseries 129} (2022) 091802},
  \href{http://arxiv.org/abs/2204.03693}{[{\ttfamily 2204.03693}]}.

\bibitem{Zhu:2022tpr}
C.-R. Zhu, M.-Y. Cui, Z.-Q. Xia, Z.-H. Yu, X.~Huang, Q.~Yuan, and Y.-Z. Fan,
  ``{Explaining the GeV Antiproton Excess, GeV \ensuremath{\gamma}-Ray Excess,
  and W-Boson Mass Anomaly in an Inert Two Higgs Doublet Model},''
  \href{http://dx.doi.org/10.1103/PhysRevLett.129.231101}{{\em Phys. Rev.
  Lett.} {\bfseries 129} (2022) 231101},
  \href{http://arxiv.org/abs/2204.03767}{[{\ttfamily 2204.03767}]}.

\bibitem{Ilnicka:2015jba}
A.~Ilnicka, M.~Krawczyk, and T.~Robens, ``{Inert Doublet Model in light of LHC
  Run I and astrophysical data},''
  \href{http://dx.doi.org/10.1103/PhysRevD.93.055026}{{\em Phys. Rev. D}
  {\bfseries 93} (2016) 055026},
  \href{http://arxiv.org/abs/1508.01671}{[{\ttfamily 1508.01671}]}.

\bibitem{Branco:2011iw}
G.~C. Branco, P.~M. Ferreira, L.~Lavoura, M.~N. Rebelo, M.~Sher, and J.~P.
  Silva, ``{Theory and phenomenology of two-Higgs-doublet models},''
  \href{http://dx.doi.org/10.1016/j.physrep.2012.02.002}{{\em Phys. Rept.}
  {\bfseries 516} (2012) 1--102},
  \href{http://arxiv.org/abs/1106.0034}{[{\ttfamily 1106.0034}]}.

\bibitem{Belyaev:2016lok}
A.~Belyaev, G.~Cacciapaglia, I.~P. Ivanov, F.~Rojas-Abatte, and M.~Thomas,
  ``{Anatomy of the Inert Two Higgs Doublet Model in the light of the LHC and
  non-LHC Dark Matter Searches},''
  \href{http://dx.doi.org/10.1103/PhysRevD.97.035011}{{\em Phys. Rev. D}
  {\bfseries 97} (2018) 035011},
  \href{http://arxiv.org/abs/1612.00511}{[{\ttfamily 1612.00511}]}.

\bibitem{Workman:2022ynf}
{\bfseries Particle Data Group} Collaboration, R.~L. Workman {\em et~al.},
  ``{Review of Particle Physics},''
  \href{http://dx.doi.org/10.1093/ptep/ptac097}{{\em PTEP} {\bfseries 2022}
  (2022) 083C01}.

\bibitem{Kannike:2012pe}
K.~Kannike, ``{Vacuum Stability Conditions From Copositivity Criteria},''
  \href{http://dx.doi.org/10.1140/epjc/s10052-012-2093-z}{{\em Eur. Phys. J. C}
  {\bfseries 72} (2012) 2093},
  \href{http://arxiv.org/abs/1205.3781}{[{\ttfamily 1205.3781}]}.

\bibitem{Merle:2015ica}
A.~Merle and M.~Platscher, ``{Running of radiative neutrino masses: the
  scotogenic model \textemdash{} revisited},''
  \href{http://dx.doi.org/10.1007/JHEP11(2015)148}{{\em JHEP} {\bfseries 11}
  (2015) 148}, \href{http://arxiv.org/abs/1507.06314}{[{\ttfamily
  1507.06314}]}.

\bibitem{tHooft:1979rat}
G.~'t~Hooft, ``{Naturalness, chiral symmetry, and spontaneous chiral symmetry
  breaking},'' \href{http://dx.doi.org/10.1007/978-1-4684-7571-5_9}{{\em NATO
  Sci. Ser. B} {\bfseries 59} (1980) 135--157}.

\bibitem{Casas:2001sr}
J.~A. Casas and A.~Ibarra, ``{Oscillating neutrinos and $\mu \to e, \gamma$},''
  \href{http://dx.doi.org/10.1016/S0550-3213(01)00475-8}{{\em Nucl. Phys. B}
  {\bfseries 618} (2001) 171--204},
  \href{http://arxiv.org/abs/hep-ph/0103065}{[{\ttfamily hep-ph/0103065}]}.

\bibitem{Toma:2013zsa}
T.~Toma and A.~Vicente, ``{Lepton Flavor Violation in the Scotogenic Model},''
  \href{http://dx.doi.org/10.1007/JHEP01(2014)160}{{\em JHEP} {\bfseries 01}
  (2014) 160}, \href{http://arxiv.org/abs/1312.2840}{[{\ttfamily 1312.2840}]}.

\bibitem{Esteban:2020cvm}
I.~Esteban, M.~C. Gonzalez-Garcia, M.~Maltoni, T.~Schwetz, and A.~Zhou, ``{The
  fate of hints: updated global analysis of three-flavor neutrino
  oscillations},'' \href{http://dx.doi.org/10.1007/JHEP09(2020)178}{{\em JHEP}
  {\bfseries 09} (2020) 178},
  \href{http://arxiv.org/abs/2007.14792}{[{\ttfamily 2007.14792}]}. NuFit 5.1
  (2021) from \url{www.nu-fit.org}.

\bibitem{Planck:2018vyg}
{\bfseries Planck} Collaboration, N.~Aghanim {\em et~al.}, ``{Planck 2018
  results. VI. Cosmological parameters},''
  \href{http://dx.doi.org/10.1051/0004-6361/201833910}{{\em Astron. Astrophys.}
  {\bfseries 641} (2020) A6},
  \href{http://arxiv.org/abs/1807.06209}{[{\ttfamily 1807.06209}]}. [Erratum:
  Astron.Astrophys. 652, C4 (2021)].

\bibitem{DiValentino:2021hoh}
E.~Di~Valentino, S.~Gariazzo, and O.~Mena, ``{Most constraining cosmological
  neutrino mass bounds},''
  \href{http://dx.doi.org/10.1103/PhysRevD.104.083504}{{\em Phys. Rev. D}
  {\bfseries 104} (2021) 083504},
  \href{http://arxiv.org/abs/2106.15267}{[{\ttfamily 2106.15267}]}.

\bibitem{Peskin:1990zt}
M.~E. Peskin and T.~Takeuchi, ``{A New constraint on a strongly interacting
  Higgs sector},'' \href{http://dx.doi.org/10.1103/PhysRevLett.65.964}{{\em
  Phys. Rev. Lett.} {\bfseries 65} (1990) 964--967}.

\bibitem{Peskin:1991sw}
M.~E. Peskin and T.~Takeuchi, ``{Estimation of oblique electroweak
  corrections},'' \href{http://dx.doi.org/10.1103/PhysRevD.46.381}{{\em Phys.
  Rev. D} {\bfseries 46} (1992) 381--409}.

\bibitem{Maksymyk:1993zm}
I.~Maksymyk, C.~P. Burgess, and D.~London, ``{Beyond S, T and U},''
  \href{http://dx.doi.org/10.1103/PhysRevD.50.529}{{\em Phys. Rev. D}
  {\bfseries 50} (1994) 529--535},
  \href{http://arxiv.org/abs/hep-ph/9306267}{[{\ttfamily hep-ph/9306267}]}.

\bibitem{Asadi:2022xiy}
P.~Asadi, C.~Cesarotti, K.~Fraser, S.~Homiller, and A.~Parikh, ``{Oblique
  Lessons from the $W$ Mass Measurement at CDF II},''
  \href{http://arxiv.org/abs/2204.05283}{[{\ttfamily 2204.05283}]}.

\bibitem{Lu:2022bgw}
C.-T. Lu, L.~Wu, Y.~Wu, and B.~Zhu, ``{Electroweak precision fit and new
  physics in light of the W boson mass},''
  \href{http://dx.doi.org/10.1103/PhysRevD.106.035034}{{\em Phys. Rev. D}
  {\bfseries 106} (2022) 035034},
  \href{http://arxiv.org/abs/2204.03796}{[{\ttfamily 2204.03796}]}.

\bibitem{Grimus:2008nb}
W.~Grimus, L.~Lavoura, O.~M. Ogreid, and P.~Osland, ``{The Oblique parameters
  in multi-Higgs-doublet models},''
  \href{http://dx.doi.org/10.1016/j.nuclphysb.2008.04.019}{{\em Nucl. Phys. B}
  {\bfseries 801} (2008) 81--96},
  \href{http://arxiv.org/abs/0802.4353}{[{\ttfamily 0802.4353}]}.

\bibitem{OPAL:2003wxm}
{\bfseries OPAL} Collaboration, G.~Abbiendi {\em et~al.}, ``{Search for
  chargino and neutralino production at s**(1/2) = 192-GeV to 209 GeV at
  LEP},'' \href{http://dx.doi.org/10.1140/epjc/s2004-01758-8}{{\em Eur. Phys.
  J. C} {\bfseries 35} (2004) 1--20},
  \href{http://arxiv.org/abs/hep-ex/0401026}{[{\ttfamily hep-ex/0401026}]}.

\bibitem{OPAL:2003nhx}
{\bfseries OPAL} Collaboration, G.~Abbiendi {\em et~al.}, ``{Search for
  anomalous production of dilepton events with missing transverse momentum in
  e+ e- collisions at s**(1/2) = 183-Gev to 209-GeV},''
  \href{http://dx.doi.org/10.1140/epjc/s2003-01466-y}{{\em Eur. Phys. J. C}
  {\bfseries 32} (2004) 453--473},
  \href{http://arxiv.org/abs/hep-ex/0309014}{[{\ttfamily hep-ex/0309014}]}.

\bibitem{ATLAS:2014hep}
{\bfseries ATLAS} Collaboration, G.~Aad {\em et~al.}, ``{Search for the direct
  production of charginos, neutralinos and staus in final states with at least
  two hadronically decaying taus and missing transverse momentum in $pp$
  collisions at $\sqrt{s}$ = 8 TeV with the ATLAS detector},''
  \href{http://dx.doi.org/10.1007/JHEP10(2014)096}{{\em JHEP} {\bfseries 10}
  (2014) 096}, \href{http://arxiv.org/abs/1407.0350}{[{\ttfamily 1407.0350}]}.

\bibitem{ATLAS:2019lff}
{\bfseries ATLAS} Collaboration, G.~Aad {\em et~al.}, ``{Search for electroweak
  production of charginos and sleptons decaying into final states with two
  leptons and missing transverse momentum in $\sqrt{s}=13$ TeV $pp$ collisions
  using the ATLAS detector},''
  \href{http://dx.doi.org/10.1140/epjc/s10052-019-7594-6}{{\em Eur. Phys. J. C}
  {\bfseries 80} (2020) 123},
  \href{http://arxiv.org/abs/1908.08215}{[{\ttfamily 1908.08215}]}.

\bibitem{ATLAS:2019gti}
{\bfseries ATLAS} Collaboration, G.~Aad {\em et~al.}, ``{Search for direct stau
  production in events with two hadronic $\tau$-leptons in $\sqrt{s} = 13$ TeV
  $pp$ collisions with the ATLAS detector},''
  \href{http://dx.doi.org/10.1103/PhysRevD.101.032009}{{\em Phys. Rev. D}
  {\bfseries 101} (2020) 032009},
  \href{http://arxiv.org/abs/1911.06660}{[{\ttfamily 1911.06660}]}.

\bibitem{CMS:2018eqb}
{\bfseries CMS} Collaboration, A.~M. Sirunyan {\em et~al.}, ``{Search for
  supersymmetric partners of electrons and muons in proton-proton collisions at
  $\sqrt{s}=$ 13 TeV},''
  \href{http://dx.doi.org/10.1016/j.physletb.2019.01.005}{{\em Phys. Lett. B}
  {\bfseries 790} (2019) 140--166},
  \href{http://arxiv.org/abs/1806.05264}{[{\ttfamily 1806.05264}]}.

\bibitem{CMS:2018yan}
{\bfseries CMS} Collaboration, A.~M. Sirunyan {\em et~al.}, ``{Search for
  supersymmetry in events with a $\tau$ lepton pair and missing transverse
  momentum in proton-proton collisions at $\sqrt{s} =$ 13 TeV},''
  \href{http://dx.doi.org/10.1007/JHEP11(2018)151}{{\em JHEP} {\bfseries 11}
  (2018) 151}, \href{http://arxiv.org/abs/1807.02048}{[{\ttfamily
  1807.02048}]}.

\bibitem{CMS:2020bfa}
{\bfseries CMS} Collaboration, A.~M. Sirunyan {\em et~al.}, ``{Search for
  supersymmetry in final states with two oppositely charged same-flavor leptons
  and missing transverse momentum in proton-proton collisions at $\sqrt{s} =$
  13 TeV},'' \href{http://dx.doi.org/10.1007/JHEP04(2021)123}{{\em JHEP}
  {\bfseries 04} (2021) 123},
  \href{http://arxiv.org/abs/2012.08600}{[{\ttfamily 2012.08600}]}.

\bibitem{Edsjo:1997bg}
J.~Edsjo and P.~Gondolo, ``{Neutralino relic density including
  coannihilations},'' \href{http://dx.doi.org/10.1103/PhysRevD.56.1879}{{\em
  Phys. Rev. D} {\bfseries 56} (1997) 1879--1894},
  \href{http://arxiv.org/abs/hep-ph/9704361}{[{\ttfamily hep-ph/9704361}]}.

\bibitem{Belanger:2014vza}
G.~B\'elanger, F.~Boudjema, A.~Pukhov, and A.~Semenov, ``{micrOMEGAs4.1: two
  dark matter candidates},''
  \href{http://dx.doi.org/10.1016/j.cpc.2015.03.003}{{\em Comput. Phys.
  Commun.} {\bfseries 192} (2015) 322--329},
  \href{http://arxiv.org/abs/1407.6129}{[{\ttfamily 1407.6129}]}.

\bibitem{Pukhov:2004ca}
A.~Pukhov, ``{CalcHEP 2.3: MSSM, structure functions, event generation, batchs,
  and generation of matrix elements for other packages},''
  \href{http://arxiv.org/abs/hep-ph/0412191}{[{\ttfamily hep-ph/0412191}]}.

\bibitem{LopezHonorez:2010eeh}
L.~Lopez~Honorez and C.~E. Yaguna, ``{The inert doublet model of dark matter
  revisited},'' \href{http://dx.doi.org/10.1007/JHEP09(2010)046}{{\em JHEP}
  {\bfseries 09} (2010) 046}, \href{http://arxiv.org/abs/1003.3125}{[{\ttfamily
  1003.3125}]}.

\bibitem{Alwall:2014hca}
J.~Alwall, R.~Frederix, S.~Frixione, V.~Hirschi, F.~Maltoni, O.~Mattelaer,
  H.~S. Shao, T.~Stelzer, P.~Torrielli, and M.~Zaro, ``{The automated
  computation of tree-level and next-to-leading order differential cross
  sections, and their matching to parton shower simulations},''
  \href{http://dx.doi.org/10.1007/JHEP07(2014)079}{{\em JHEP} {\bfseries 07}
  (2014) 079}, \href{http://arxiv.org/abs/1405.0301}{[{\ttfamily 1405.0301}]}.

\bibitem{Ambrogi:2018jqj}
F.~Ambrogi, C.~Arina, M.~Backovic, J.~Heisig, F.~Maltoni, L.~Mantani,
  O.~Mattelaer, and G.~Mohlabeng, ``{MadDM v.3.0: a Comprehensive Tool for Dark
  Matter Studies},'' \href{http://dx.doi.org/10.1016/j.dark.2018.11.009}{{\em
  Phys. Dark Univ.} {\bfseries 24} (2019) 100249},
  \href{http://arxiv.org/abs/1804.00044}{[{\ttfamily 1804.00044}]}.

\bibitem{Alguero:2022inz}
G.~Alguero, G.~Belanger, S.~Kraml, and A.~Pukhov, ``{Co-scattering in
  micrOMEGAs: A case study for the singlet-triplet dark matter model},''
  \href{http://dx.doi.org/10.21468/SciPostPhys.13.6.124}{{\em SciPost Phys.}
  {\bfseries 13} no.~6, (2022) 124},
  \href{http://arxiv.org/abs/2207.10536}{[{\ttfamily 2207.10536}]}.

\bibitem{Staub:2008uz}
F.~Staub, ``{SARAH},'' \href{http://arxiv.org/abs/0806.0538}{[{\ttfamily
  0806.0538}]}.

\bibitem{Staub:2009bi}
F.~Staub, ``{From Superpotential to Model Files for FeynArts and
  CalcHep/CompHep},'' \href{http://dx.doi.org/10.1016/j.cpc.2010.01.011}{{\em
  Comput. Phys. Commun.} {\bfseries 181} (2010) 1077--1086},
  \href{http://arxiv.org/abs/0909.2863}{[{\ttfamily 0909.2863}]}.

\bibitem{Garny:2021qsr}
M.~Garny and J.~Heisig, ``{Bound-state effects on dark matter coannihilation:
  Pushing the boundaries of conversion-driven freeze-out},''
  \href{http://dx.doi.org/10.1103/PhysRevD.105.055004}{{\em Phys. Rev. D}
  {\bfseries 105} (2022) 055004},
  \href{http://arxiv.org/abs/2112.01499}{[{\ttfamily 2112.01499}]}.

\bibitem{Lundstrom:2008ai}
E.~Lundstrom, M.~Gustafsson, and J.~Edsjo, ``{The Inert Doublet Model and LEP
  II Limits},'' \href{http://dx.doi.org/10.1103/PhysRevD.79.035013}{{\em Phys.
  Rev. D} {\bfseries 79} (2009) 035013},
  \href{http://arxiv.org/abs/0810.3924}{[{\ttfamily 0810.3924}]}.

\bibitem{LopezHonorez:2010tb}
L.~Lopez~Honorez and C.~E. Yaguna, ``{A new viable region of the inert doublet
  model},'' \href{http://dx.doi.org/10.1088/1475-7516/2011/01/002}{{\em JCAP}
  {\bfseries 01} (2011) 002}, \href{http://arxiv.org/abs/1011.1411}{[{\ttfamily
  1011.1411}]}.

\bibitem{Eiteneuer:2017hoh}
B.~Eiteneuer, A.~Goudelis, and J.~Heisig, ``{The inert doublet model in the
  light of Fermi-LAT gamma-ray data: a global fit analysis},''
  \href{http://dx.doi.org/10.1140/epjc/s10052-017-5166-1}{{\em Eur. Phys. J. C}
  {\bfseries 77} no.~9, (2017) 624},
  \href{http://arxiv.org/abs/1705.01458}{[{\ttfamily 1705.01458}]}.

\bibitem{Alguero:2021dig}
G.~Alguero, J.~Heisig, C.~K. Khosa, S.~Kraml, S.~Kulkarni, A.~Lessa,
  H.~Reyes-Gonz\'alez, W.~Waltenberger, and A.~Wongel, ``{Constraining new
  physics with SModelS version 2},''
  \href{http://dx.doi.org/10.1007/JHEP08(2022)068}{{\em JHEP} {\bfseries 08}
  (2022) 068}, \href{http://arxiv.org/abs/2112.00769}{[{\ttfamily
  2112.00769}]}.

\bibitem{Barducci:2016pcb}
D.~Barducci, G.~Belanger, J.~Bernon, F.~Boudjema, J.~Da~Silva, S.~Kraml,
  U.~Laa, and A.~Pukhov, ``{Collider limits on new physics within
  micrOMEGAs$\_$4.3},'' \href{http://dx.doi.org/10.1016/j.cpc.2017.08.028}{{\em
  Comput. Phys. Commun.} {\bfseries 222} (2018) 327--338},
  \href{http://arxiv.org/abs/1606.03834}{[{\ttfamily 1606.03834}]}.

\bibitem{ATLAS:2021moa}
{\bfseries ATLAS} Collaboration, G.~Aad {\em et~al.}, ``{Search for
  chargino\textendash{}neutralino pair production in final states with three
  leptons and missing transverse momentum in $\sqrt{s} = 13$~TeV pp collisions
  with the ATLAS detector},''
  \href{http://dx.doi.org/10.1140/epjc/s10052-021-09749-7}{{\em Eur. Phys. J.
  C} {\bfseries 81} (2021) 1118},
  \href{http://arxiv.org/abs/2106.01676}{[{\ttfamily 2106.01676}]}.

\bibitem{ATLAS:2019wgx}
{\bfseries ATLAS} Collaboration, G.~Aad {\em et~al.}, ``{Search for
  chargino-neutralino production with mass splittings near the electroweak
  scale in three-lepton final states in $\sqrt {s}$=13 TeV $pp$ collisions with
  the ATLAS detector},''
  \href{http://dx.doi.org/10.1103/PhysRevD.101.072001}{{\em Phys. Rev. D}
  {\bfseries 101} (2020) 072001},
  \href{http://arxiv.org/abs/1912.08479}{[{\ttfamily 1912.08479}]}.

\bibitem{CMS:2021edw}
{\bfseries CMS} Collaboration, A.~Tumasyan {\em et~al.}, ``{Search for
  supersymmetry in final states with two or three soft leptons and missing
  transverse momentum in proton-proton collisions at $ \sqrt{s} $ = 13 TeV},''
  \href{http://dx.doi.org/10.1007/JHEP04(2022)091}{{\em JHEP} {\bfseries 04}
  (2022) 091}, \href{http://arxiv.org/abs/2111.06296}{[{\ttfamily
  2111.06296}]}.

\bibitem{CMS:2017moi}
{\bfseries CMS} Collaboration, A.~M. Sirunyan {\em et~al.}, ``{Search for
  electroweak production of charginos and neutralinos in multilepton final
  states in proton-proton collisions at $\sqrt{s}=$ 13 TeV},''
  \href{http://dx.doi.org/10.1007/JHEP03(2018)166}{{\em JHEP} {\bfseries 03}
  (2018) 166}, \href{http://arxiv.org/abs/1709.05406}{[{\ttfamily
  1709.05406}]}.

\bibitem{ATLAS:2019gqq}
{\bfseries ATLAS} Collaboration, M.~Aaboud {\em et~al.}, ``{Search for heavy
  charged long-lived particles in the ATLAS detector in 36.1 fb$^{-1}$ of
  proton-proton collision data at $\sqrt{s} = 13$ TeV},''
  \href{http://dx.doi.org/10.1103/PhysRevD.99.092007}{{\em Phys. Rev. D}
  {\bfseries 99} (2019) 092007},
  \href{http://arxiv.org/abs/1902.01636}{[{\ttfamily 1902.01636}]}.

\bibitem{CMS:2015lsu}
{\bfseries CMS} Collaboration, V.~Khachatryan {\em et~al.}, ``{Constraints on
  the pMSSM, AMSB model and on other models from the search for long-lived
  charged particles in proton-proton collisions at sqrt(s) = 8 TeV},''
  \href{http://dx.doi.org/10.1140/epjc/s10052-015-3533-3}{{\em Eur. Phys. J. C}
  {\bfseries 75} (2015) 325},
  \href{http://arxiv.org/abs/1502.02522}{[{\ttfamily 1502.02522}]}.

\bibitem{CMS:2013czn}
{\bfseries CMS} Collaboration, S.~Chatrchyan {\em et~al.}, ``{Searches for
  Long-Lived Charged Particles in $pp$ Collisions at $\sqrt{s}$=7 and 8 TeV},''
  \href{http://dx.doi.org/10.1007/JHEP07(2013)122}{{\em JHEP} {\bfseries 07}
  (2013) 122}, \href{http://arxiv.org/abs/1305.0491}{[{\ttfamily 1305.0491}]}.

\bibitem{ATLAS:2017oal}
{\bfseries ATLAS} Collaboration, M.~Aaboud {\em et~al.}, ``{Search for
  long-lived charginos based on a disappearing-track signature in pp collisions
  at $ \sqrt{s}=13 $ TeV with the ATLAS detector},''
  \href{http://dx.doi.org/10.1007/JHEP06(2018)022}{{\em JHEP} {\bfseries 06}
  (2018) 022}, \href{http://arxiv.org/abs/1712.02118}{[{\ttfamily
  1712.02118}]}.

\bibitem{ATLAS:2020wjh}
{\bfseries ATLAS} Collaboration, G.~Aad {\em et~al.}, ``{Search for Displaced
  Leptons in $\sqrt{s} = 13$ TeV $pp$ Collisions with the ATLAS Detector},''
  \href{http://dx.doi.org/10.1103/PhysRevLett.127.051802}{{\em Phys. Rev.
  Lett.} {\bfseries 127} (2021) 051802},
  \href{http://arxiv.org/abs/2011.07812}{[{\ttfamily 2011.07812}]}.

\bibitem{Rodejohann:2011mu}
W.~Rodejohann, ``{Neutrino-less Double Beta Decay and Particle Physics},''
  \href{http://dx.doi.org/10.1142/S0218301311020186}{{\em Int. J. Mod. Phys. E}
  {\bfseries 20} (2011) 1833--1930},
  \href{http://arxiv.org/abs/1106.1334}{[{\ttfamily 1106.1334}]}.

\end{thebibliography}\endgroup

\end{document}